\documentclass[llinenumbers, trackchanges, twocolumn]{aastex631}
\usepackage{latexsym}
\usepackage{graphicx}
\usepackage{color}
\usepackage{amsmath}
\usepackage{comment}
\usepackage{hyperref}
\usepackage{bm}
\usepackage{acronym}
\usepackage{url}
\usepackage{textcomp}
\usepackage{xspace}
\usepackage{acronym}

\newcommand{\code}[1]{\texttt{#1}\xspace}

\newcommand{\Ntrial}{N_{\mathrm{templ}}}
\newcommand{\Tobs}{T_{\mathrm{obs}}}
\newcommand{\pfa}{p_{\rm fa}}

\setcitestyle{citesep={, }}

\begin{document}

\title{Narrowband searches for continuous gravitational waves from known pulsars in the first two parts of the fourth LIGO--Virgo--KAGRA observing run}


\author[0000-0003-4786-2698]{A.~G.~Abac}
\affiliation{Max Planck Institute for Gravitational Physics (Albert Einstein Institute), D-14476 Potsdam, Germany}
\author{I.~Abouelfettouh}
\affiliation{LIGO Hanford Observatory, Richland, WA 99352, USA}
\author{F.~Acernese}
\affiliation{Dipartimento di Farmacia, Universit\`a di Salerno, I-84084 Fisciano, Salerno, Italy}
\affiliation{INFN, Sezione di Napoli, I-80126 Napoli, Italy}
\author[0000-0002-8648-0767]{K.~Ackley}
\affiliation{University of Warwick, Coventry CV4 7AL, United Kingdom}
\author{A.~Adam}
\affiliation{OzGrav, University of Western Australia, Crawley, Western Australia 6009, Australia}
\author[0000-0001-5525-6255]{C.~Adamcewicz}
\affiliation{OzGrav, School of Physics \& Astronomy, Monash University, Clayton 3800, Victoria, Australia}
\author[0009-0004-2101-5428]{S.~Adhicary}
\affiliation{The Pennsylvania State University, University Park, PA 16802, USA}
\author{D.~Adhikari}
\affiliation{Max Planck Institute for Gravitational Physics (Albert Einstein Institute), D-30167 Hannover, Germany}
\affiliation{Leibniz Universit\"{a}t Hannover, D-30167 Hannover, Germany}
\author[0000-0002-4559-8427]{N.~Adhikari}
\affiliation{University of Wisconsin-Milwaukee, Milwaukee, WI 53201, USA}
\author[0000-0002-5731-5076]{R.~X.~Adhikari}
\affiliation{LIGO Laboratory, California Institute of Technology, Pasadena, CA 91125, USA}
% ...

\collaboration{3000}{The LIGO Scientific Collaboration, the Virgo Collaboration, and the KAGRA Collaboration}


\newcommand{\JBO}{Jodrell Bank Centre for Astrophysics, School of Physics and Astronomy, University of Manchester, Manchester, UK, M13 9PL}
\newcommand{\IAR}{Instituto Argentino de Radioastronomía (CCT La Plata, CONICET; CICPBA; UNLP), C.C.5, (1894) Villa Elisa, Buenos Aires, Argentina}
\newcommand{\FacCienAstr}{Facultad de Ciencias Astronómicas y Geofísicas, Universidad Nacional de La Plata, Paseo del Bosque, B1900FWA La Plata, Argentina}
\newcommand{\XRayAstrophysicsLaboratory}{X-Ray Astrophysics Laboratory, NASA Goddard Space Flight Center, Greenbelt, MD, 20771, USA}
\newcommand{\ozgrav}{ARC Center of Excellence for Gravitational Wave Discovery (OzGrav), Swinburne University of Technology, VIC 3122, Australia}
\newcommand{\Swinburne}{Centre for Astrophysics and Supercomputing, Swinburne University of Technology, Hawthorn, VIC 3122, Australia}
\newcommand{\CNRS}{LPC2E, OSUC, Univ Orleans, CNRS, CNES, Observatoire de Paris, F-45071 Orleans, France}
\newcommand{\ORN}{ORN, Observatoire de Paris, Universit\'e PSL, Univ Orl\'eans, CNRS, 18330 Nan\c{c}ay, France}

\author[0000-0003-4027-4826]{S.B.~Araujo Furlan}
\affiliation{\IAR}

\author[0009-0008-6187-8753]{Z. Arzoumanian}
\affiliation{\XRayAstrophysicsLaboratory}

\author[0000-0003-3265-2866]{E. Carli}
\affiliation{\Swinburne}\affiliation{\ozgrav}

\author[0000-0002-1775-9692]{I.~Cognard}
\affiliation{\CNRS}\affiliation{\ORN}

\author[0000-0002-7031-4828]{M.~Cury\l{}o}
\affiliation{School of Physics and Astronomy, Monash University, Clayton VIC 3800, Australia}\affiliation{\ozgrav}

\author[0000-0002-5761-2417]{S. del Palacio}
\affiliation{\IAR}
\affiliation{Department of Space, Earth and Environment, Chalmers University of Technology, SE-412 96 Gothenburg, Sweden}

\author[0000-0003-2481-2348]{C. M. Espinoza}
\affiliation{Departamento de F\'isica, Universidad de Santiago de Chile (USACH), Av. V\'ictor Jara 3493, Estaci\'on Central, Chile}
\affiliation{Center for Interdisciplinary Research in Astrophysics and Space Sciences (CIRAS), Universidad de Santiago de Chile, Santiago, Chile}

\author[0000-0001-8384-5049]{E.~Fonseca}
\affiliation{Department of Physics and Astronomy, West Virginia University, PO Box 6315, Morgantown, WV 26506, USA}
\affiliation{Center for Gravitational Waves and Cosmology, West Virginia University, Chestnut Ridge Research Building, Morgantown, WV 26505, USA}

\author[0000-0003-1282-3031]{G. Gancio}
\affiliation{\IAR}

\author[0000-0001-9072-4069]{F. Garc\'{i}a}
\affiliation{\IAR}
\affiliation{\FacCienAstr}

\author[0000-0001-7115-2819]{K. C. Gendreau}
\affiliation{\XRayAstrophysicsLaboratory}

\author[0000-0002-9049-8716]{L.~Guillemot}
\affiliation{\CNRS}\affiliation{\ORN}

\author[0000-0002-6449-106X]{S. Guillot}
\affiliation{IRAP, CNRS, 9 avenue du Colonel Roche, BP 44346, 31028 Toulouse Cedex 4, France}
\affiliation{Universit\'e de Toulouse, CNES, UPS-OMP, 31028 Toulouse, France}

\author[0000-0001-5567-5492]{M. J. Keith}
\affiliation{\JBO}

\author[0000-0002-7889-6586]{L. Kuiper}
\affiliation{Space Research Organisation Netherlands, Niels Bohrweg 4, 2333 CA, Leiden, Netherlands}

\author[0000-0002-4799-1281]{A. G. Lyne}
\affiliation{\JBO}

\author[0000-0001-8845-1225]{B.~W.~Meyers}\affiliation{International Centre for Radio Astronomy Research, Curtin University, Bentley, WA, Australia}\affiliation{Australian SKA Regional Centre (AusSRC), Curtin University, Bentley, WA, Australia}

\author[0000-0002-5455-3474]{M.~T.~Miles}
\affiliation{Department of Physics and Astronomy, Vanderbilt University, 2301 Vanderbilt Place, Nashville, TN 37235, USA}\affiliation{\ozgrav}

\author[0000-0001-8691-8039]{J.~L.~Palfreyman}
\affiliation{School of Natural Sciences, University of Tasmania, Hobart, Australia}

\author[0000-0002-8912-0732]{A.~B.~Pearlman}
\affiliation{Department of Physics, McGill University, 3600 rue University, Montréal, QC H3A 2T8, Canada}
\affiliation{Trottier Space Institute, McGill University, 3550 rue University, Montréal, QC H3A 2A7, Canada}

\author[0000-0002-2035-4688]{D.~J.~Reardon}
\affiliation{\Swinburne}\affiliation{\ozgrav}

\author[0000-0002-5260-1807]{G.~E.~Romero}
\affiliation{\IAR}
\affiliation{\FacCienAstr}

\author[0000-0002-7285-6348]{R.~M.~Shannon}
\affiliation{\Swinburne}\affiliation{\ozgrav}

\author[0000-0002-9581-2452]{B. Shaw}
\affiliation{\JBO}

\author[0000-0001-9784-8670]{I.~H.~ Stairs}
\affiliation{Department of Physics and Astronomy, University of British Columbia, 6224 Agricultural Road, Vancouver, BC V6T 1Z1, Canada}

\author[0000-0001-9242-7041]{B. W. Stappers}
\affiliation{\JBO}

\author[0000-0002-3649-276X]{G.~Theureau}
\affiliation{\CNRS}\affiliation{\ORN}

\author[0000-0003-2122-4540]{P. Weltevrede}
\affiliation{\JBO}

\author[0009-0009-5593-367X]{E.~Zubieta}
\affiliation{\IAR}
\affiliation{\FacCienAstr}

\begin{abstract}
Rotating nonaxisymmetric neutron stars (NSs) are promising sources for continuous gravitational waves (CWs). CWs may, if detected, inform us about the internal structure and equation of state of NSs. Here, we present a narrowband search for CWs from known pulsars, for which a matched-filter search can be applied. Narrowband searches are robust to mismatches between electromagnetic (EM) and gravitational emissions, in contrast to fully targeted searches where they are assumed to be phase-locked. In this work, we search for the CW counterparts emitted by 34 pulsars using data from the first and second parts of the fourth LIGO--Virgo--KAGRA observing run. We use the 5n-vector narrowband pipeline, which applies frequency-domain matched filtering. In previous searches, it covered a narrow range in the frequency -- frequency time derivative ($f$ -- $\dot{f}$) space. Here, we also explore a range in the second time derivative of the frequency $\ddot{f}$ around the EM observations. Additionally, for the first time, we target sources in a binary system with this kind of search. We find no evidence for CWs and therefore set upper limits on the strain amplitude emitted by each pulsar. For 20 analyses, we report an upper limit below the theoretical spin-down limit. The tightest constraint is for pulsar PSR J0534+2200 (the Crab pulsar), for which our strain upper limit on the CW amplitude corresponds to $\leq 0.04\%$ of the spin-down power being radiated in the CW channel.
\end{abstract}

\section{Introduction}
Continuous gravitational waves (CWs) represent one of the most elusive and scientifically rich targets in gravitational wave (GW) astronomy. Unlike transient signals produced by cataclysmic events such as binary mergers of black holes and/or neutron stars (NSs), CWs are expected to be persistent and nearly monochromatic signals potentially emitted by rapidly rotating NSs with a nonsymmetric mass distribution with respect to the rotation axis \citep{2023LRR....26....3R}. 

The mass-distribution asymmetries, such as ``mountains'' on the crust \citep{Gittins_2024} or internal deformations supported by strong magnetic fields \citep{1996AA...312..675B, 2002PhRvD..66h4025C} or exotic matter \citep{2005PhRvL..95u1101O}, can cause the star to emit gravitational radiation steadily over timescales of months to years. In the literature \citep[see][]{Haskell2023-fu, 2023LRR....26....3R, WETTE2023102880}, different emission models are generally considered; the single harmonic model predicts an emission at exactly twice the spin frequency, while the dual harmonic model has emission at both 1 and 2 times the spin frequency \citep{2010MNRAS.402.2503J}. Additional mechanisms may also contribute to CW emission, such as free precession or internal fluid oscillation modes (e.g., \textit{r}-modes) \citep{1998ApJ...502..708A, 1998ApJ...502..714F}. The search for CWs is motivated by their potential to unveil fundamental information about the structure and composition of NSs~\citep{Haskell2023-fu, Owen:2025ata}. A detection would provide direct evidence of nonaxisymmetric deformations, offer insight into the dense-matter physics of the NS interior~\citep[see e.g.,][]{Idrisy_2015, Ghosh_2023,Ghosh_MNRAS2023}, and allow tests of gravitational theories in the strong-field regime~\citep{2017PhRvD..96d2001I, PhysRevD.100.104036}. 

However, CW signals are expected to be extremely weak and much weaker than the transient signals detected so far by the LIGO--Virgo--KAGRA (LVK) Collaboration (see~\citet{2025arXiv250818082T} for the latest catalog of GW detections, and \citet{2015CQGra..32g4001L,2015CQGra..32b4001A,KAGRA:2020tym} for more information on LVK detectors). This low amplitude necessitates long observation times and sensitive data analysis techniques capable of extracting faint, long-duration signals from detector noise. Searches for CWs must also account for intrinsic decrease in the NS’s spin frequency (the so-called spin-down), and the Doppler modulation caused by Earth’s motion and by the orbital motion for sources in binary systems.

To address these challenges, different types of CW search strategies have been developed \citep{2023LRR....26....3R, WETTE2023102880}, each optimized for different assumptions about the signal and the level of prior information available about the source. Known pulsars for which we have accurate measurements of their sky positions and rotation parameters from electromagnetic (EM) observations can be targeted using fully coherent searches. The most sensitive CW searches assume that the GW phase evolution is locked to the EM phase evolution and remains consistent throughout the entire observation time, performing fully coherent analyses with a single frequency evolution template per target.

Coherent narrowband searches, the focus of this paper, are also usually used for pulsars with known position, but when the signal frequency, while known, is uncertain or may have shifted from EM predictions~\citep[e.g.,][]{2008ApJ...683L..45A, LIGOScientific:2014uzs, 2017PhRvD..96l2006A, O2Narrowband, LIGOScientific:2021quq, Ashok:2021fnj, LIGOScientific:2025kei}\footnote{Broader searches, either for sources at known positions but with unknown frequencies or over the full sky, usually use semicoherent methods, see, e.g. \citet{WETTE2023102880} for an overview.}. This could happen due to timing noise, glitches, or mismatches between gravitational and EM emission mechanisms~\citep{Ashton:2017wui,2025PASA...42..118A}. Narrowband searches extend the targeted approach by scanning a small frequency band and a limited range of spin-down parameters, achieving improved robustness while maintaining relatively high sensitivity.
Pulsars, NSs possessing powerful magnetic fields that emit beams of EM waves across different frequency bands (radio, X-rays, gamma rays), are the primary candidates for CW targeted and fully coherent narrowband searches.

In this work, we present a narrowband CW search using data from the recent LVK O4 observing run, considering the first two parts, namely, O4a and O4b data from the LIGO Livingston (L1) and LIGO Hanford detectors (H1). 
We study 34 pulsars, with ephemerides provided by a set of gamma-ray, radio, and X-ray observatories, representing the largest set of targets ever considered for a narrowband search.
For the first time, we consider sources in binary systems and explore templates in the second-order spin-down ($\ddot{f}$) dimension, enlarging the science scope of this work.

The paper is organized as follows. Section~\ref{sec:data} describes the EM and GW data used in the analysis.
Section~\ref{sec:signal_and_method} outlines the astrophysical motivations and characteristics of the expected signals and describes the analysis method.
The results are presented and discussed in Section~\ref{sec:results}. 
Our conclusions are given in Section~\ref{sec:conclusions}.

\section{Datasets\label{sec:data}}
\subsection{EM data\label{sec:em_data}}

We used EM data to obtain pulsar timing solutions that were used as inputs for the GW search. A gamma-ray timing solution was obtained for one target from Fermi Large Area Telescope~\citep[LAT;][]{2009ApJ...697.1071A}. Radio data for multiple pulsars were collected from the Nancay Radio Telescope~\citep[NRT;][]{NRT2023}, the 42 ft and Lovell telescopes at Jodrell Bank Observatory (JBO), the Argentine Institute of Radio astronomy~\citep[IAR;][]{2020A&A...633A..84G}, the Canadian Hydrogen Intensity Mapping Experiment \citep[CHIME,][]{2021ApJS..255....5C}, the MeerKAT radio telescope~\citep{Jonas:2018Jr}, and the Mount Pleasant Radio Observatory\footnote{Due to some mechanical issues in mid 2023, this observatory was not able to provide ephemerides entirely covering the O4ab run.}~\citep{Palfreyman:2011kh}.
The MeerKAT timing solutions were derived from observations over 6 yr, collected for the MeerKAT Pulsar Timing Array (MPTA). These data were processed following the methods used for the first two data releases of the MPTA \citep{Miles+23, Miles+25}.
X-ray timing solutions were obtained from Chandra~\citep{2002PASP..114....1W} and the Neutron Star Interior Composition Explorer~\citep[NICER;][]{lens.org/083-235-561-223-456}.

For processing the EM observations, we used PSRCHIVE \citep{2012AR&T....9..237V} or PRESTO \citep{2002AJ....124.1788R,2011ascl.soft07017R} packages, the Chandra Interactive Analysis of Observations (CIAO) package and Calibration Database (CALDB)~\citep{2006SPIE.6270E..1VF}, the NICER software in HEASoft~\citep{2014ascl.soft08004N}, and for the gamma-ray data, we used the procedures detailed in~\citet{Kuiper:2009fm}. We first cleaned the data for noise/radio frequency interference, when applicable. Next, we folded the observations and obtained the time of arrival (ToA) by cross-correlating the folded profiles with a template with a high signal-to-noise ratio. We selected the ToAs during the O4ab run and used TEMPO2 \citep{2006MNRAS.372.1549E, 2006MNRAS.369..655H, tempo23} or PINT \citep{2019ascl.soft02007L,2021ApJ...911...45L} to characterize the rotation of each pulsar by fitting the ToAs with a Taylor expansion of the rotational phase 
\begin{equation}\label{eq:timing-model}
    \phi(t)=\phi_0+ 2\pi,[f_{\rm rot}(t-t_0)+\frac{1}{2}\dot{f}_{\rm rot}(t-t_0)^2+\frac{1}{6}\Ddot{f}_{\rm rot}(t-t_0)^3 + ...]\,.
\end{equation}
\noindent Here, $f_{\rm rot}$, $\dot{f}_{\rm rot},$ and $\Ddot{f}_{\rm rot}$ are the rotation frequency of the pulsar, and its first and second derivatives, respectively. Their reference epoch is $t_0$, and $\phi_0$ is the phase at $t_0$. Higher-order derivatives can be included if necessary.

\subsubsection{Glitches\label{sec:glitches}}

Pulsars are known for their extremely stable rotation; however, some of them present \textit{glitches}. Glitches are abrupt changes in the rotational frequency of the pulsar, and they have been observed in over 200 pulsars~\citep{2011MNRAS.414.1679E, 2013MNRAS.429..688Y, 2022MNRAS.510.4049B}. Most glitches are likely a consequence of the interaction between
the superfluid interiors of NSs and their solid crusts. However, their dynamics and the mechanisms that may trigger them are not well understood~\citep{2022Univ....8..641Z, 2022RPPh...85l6901A}.

The additional phase in pulsar rotation induced by glitches is included in the timing model as~\citep{1987AuJPh..40..725M}:
\begin{multline}\label{eq:glitch-model}
    \phi_\mathrm{g}(t) = \Delta \phi + 2\pi\,[\Delta f_\mathrm{rot} (t-t_\mathrm{g}) + \frac{1}{2} \Delta \dot{f}_\mathrm{rot} (t-t_\mathrm{g})^2 + \\ 
    \frac{1}{6} \Delta \ddot{f}_\mathrm{rot}(t-t_\mathrm{g})^3 + \sum_i \left[1-\exp{\left(-\frac{t-t_\mathrm{g}}{\tau^{i}_\mathrm{d}}\right)} \right]\Delta f^{i,d}_\mathrm{rot} \, \tau_\mathrm{d}^{i}],
\end{multline}

\noindent where $t_\mathrm{g}$ is the glitch epoch, and $\Delta \phi$ is used to counteract its uncertainty. $\Delta f_\mathrm{rot}$, $\Delta \dot{f}_\mathrm{rot}$, and $\Delta \ddot{f}_\mathrm{rot}$ are the step changes in $f_\mathrm{rot}$, $\dot f_\mathrm{rot},$ and $\ddot{f}_\mathrm{rot}$ at $t_\mathrm{g}$, respectively. Finally, $\Delta f^{d,i}_\mathrm{rot}$ represent temporary increases in frequency that recover in $\tau_{d,i}$  days. Detected timescales for well-monitored pulsars can range from minutes 
\citep{2002ApJ...564L..85D, 2018Natur.556..219P,2019NatAs...3.1143A,2025A&A...698A..72Z} to $\sim500~\mathrm{days}$ \citep{2024A&A...689A.191Z}.

For some pulsars, we consider data even before the beginning of the run to improve the parameter estimation.
This results in the modeling of some pre-O4 glitches listed in Table~\ref{tab:glitches_O4ab}, together with the others that occurred during the run.
Details of glitching pulsar analyses will be provided in Section~\ref{sec:signal_and_method}.

\begin{table}[!ht]
    \centering
    \begin{tabular}{cc}
        \hline\hline
        PSR & $t_{\rm g}$ [MJD] \\
        \hline
        J0058\textminus7218 & 60291\\
        J0537\textminus6910 & 60223, 60379, 60611\\
        J0540\textminus6919 & 60105, 60170\\
        J0835\textminus4510 & 60430\\
        J1809\textminus1917 & 60071\\
        J1813\textminus1246 & 60247\\
        J2021+3651 & 60289\\
        J2022+3842 & 60540\\
        J2229+6114 & 60065\\
         \hline\hline
    \end{tabular}
    \caption{Pulsar glitches incorporated in the O4ab narrowband search. Details of the two PSR J0540\textminus6919 timing events are given in~\citet{Espinoza:2024cwk}. We truncated the analysis to MJD~60510 for PSR J2022+3842 due to a lack of observations to narrow down the post-glitch parameters. PSR J1809-1917 and PSR J2229+6114 glitched before the beginning of the run.}
    \label{tab:glitches_O4ab}
\end{table}

\subsection{GW data\label{sec:gw_data}}
We considered data from the first (O4a) and second (O4b) parts of the fourth observing run, known as O4ab, of the LIGO Livingston (L1) and LIGO Hanford (H1) detectors.
The first part of the run~\citep{LIGOScientific:2025snk} took place between 2023 May 24, 15:00:00 UTC (MJD 60088) and ended 2024 January 16, 16:00:00 UTC (MJD 60325). The duty factors for L1 and H1 were 69.0\% and 67.5\%, respectively.
Detectors resumed the observing mode for O4b on April 10, 2024, at 15:00 UTC (MJD 60410) and ended on 2025 January 28, at 17:00:00 UTC (MJD  60703).
Here, the duty factors for L1 and H1 were 68.1\% and 48.6\%, respectively.
As a result, we here consider $\sim 360$~days ($\sim 300$~days) for L1 (H1)  of effective observing time.

We use \textsc{L1:GDS--CALIB\_STRAIN\_CLEAN\_AR} and \textsc{H1:GDS--CALIB\_STRAIN\_CLEAN\_AR} frame channels with \textsc{CAT1} vetoes~\citep{Davis:2018yrz} for L1 and H1, respectively, following~\citet{segment_list}. For the PSR J0534+2200 search only, we substitute data from 2024 September 3 to November 11, with \textsc{L1:DCS-CALIB\_STRAIN\_CLEAN\_C01} and \textsc{H1:DCS-CALIB\_STRAIN\_CLEAN\_C01} channels since the 60~Hz line subtraction was not working well during this period in the online calibration, requiring an improved ``C01'' version~\citep{broadening_60Hz_line}.

A general description of the LIGO detectors' performance throughout O4ab is given in~\citet{LIGOO4Detector:2023wmz, membersoftheLIGOScientific:2024elc, Capote:2024rmo, LIGO:2024kkz}.
For the O4ab data used in this analysis, the worst 1$\sigma$ calibration uncertainty is within $\lesssim10$\% in amplitude and $\lesssim10^\circ$  in phase, over the range 10–2000 Hz~\citep{Wade:2025tgt}. The uncertainty at specific frequencies or times can be significantly smaller. 
Two hours of H1 data with large calibration uncertainties ($\lesssim 25\%$ in amplitude and $\lesssim15^\circ$ in phase, over the range 10–2000 Hz)~\citep{calibration_issue_H1} were inadvertently included in the analysis.
These data represent $\sim0.03\%$ of the effective observing time analyzed from the O4 run. We have confirmed, using spot checks of the hardware injections\footnote{Simulated CW signals were injected through the detector hardware for testing purposes.}~\citep{Baxi:2026wbr}, that including these data has a $<0.5\%$ effect on our results.

During the proofs stage, it was realized that the dataset used in the analyses omitted $\sim20$ days of data, between the official start of the O4b run on 2024 April 10 and the start date of the analyses on 2024 May 1. The omission of these﻿ data, which would have contributed $\sim2\%$ to the search sensitivity, does not appreciably affect the results or conclusions presented in this paper.

The Virgo and KAGRA detectors have not been considered since they joined the O4 run, only in the second and third parts, respectively.
For a description of the upgrades to the Advanced LIGO~\citep{Capote:2024rmo}, Advanced Virgo~\citep{VIRGO:2025sym}, and KAGRA detectors in preparation for the O4 run, we refer to Appendix A in~\citet{Abac_2024}.

\section{Signal model and search method~\label{sec:signal_and_method}}

In this section, we will briefly discuss the expected signal~\citep{2023LRR....26....3R} and the narrowband search method based on the five-vector formalism~\citep{2010CQGra..27s4016A,Mastrogiovanni:2017xjr} as implemented in the Snag framework~\citep{snag}.
We will focus on the single harmonic emission model since the narrowband searches are performed around $f_{\rm gw}=2f_{\rm rot}$, where $f_{\rm gw}$ is the GW frequency.
For a triaxial star, the amplitude $h_0$ can be expressed as
\begin{equation}
\label{eq:h0}
    h_0 = \frac{16\pi^2 G}{c^4} \frac{I_{zz} \varepsilon f_{\rm rot}^2 }{d} \,,
\end{equation}
where $d$ is the distance of the source, and $\varepsilon$ the equatorial ellipticity defined as 
\begin{equation}
\label{eq:epsilon_definition}
    \varepsilon \equiv \frac{|I_{xx} - I_{yy}|}{I_{zz}} \,,
\end{equation}
where $I_{xx}$, $I_{yy}$, and $I_{zz}$ are the source's principal moments of inertia, with the star rotating about the $z$-axis.

The spin-down limit $h_0^{\rm sd}$ quantifies the maximum CW amplitude assuming all the lost rotational energy is converted into GW emission.
It is calculated as
\begin{equation}
\label{eq:spin-down}
    h_0^{\rm sd}=\frac{1}{d}\left(\frac{5GI_{zz}}{2c^3}\frac{|\dot{f}_{\rm rot}|}{f_{\rm rot}}\right)^{1/2} \,.
\end{equation}
By constraining the CW emission, we can set upper bounds on the fraction of energy emitted through the gravitational channel.

The relative motion of Earth and the source modulates the signal received at the detector~\citep{WETTE2023102880}. 
It depends on the source's sky location (Earth motion) and, where needed, on up to five Keplerian parameters for the binary orbit (source motion).
After we correct the data for spin-down and Doppler effects (induced by the Earth's revolution and by the source binary motion), the expected CW signal is modulated in amplitude by the Earth's rotation and can be expressed as the real part of the following complex expression~\citep{2010CQGra..27s4016A}
\begin{equation}
h(t) = H_0 \left( H_+ \mathbf{A}^+ + H_\times \mathbf{A}^\times \right) \cdot \mathbf{W} \, e^{i(\omega_{\rm gw} t + \phi)},
\end{equation}
where $\omega_{\rm gw} = 2\pi f_{\rm gw}$, and boldface symbols denote 5-vectors (arrays with five complex components). The $\mathbf{A}^{+/\times}$ represent the five-vectors of the single-detector response to the two GW polarizations, and $\mathbf{W} = e^{i k \Theta}$ encodes sidereal modulation, with $k = \{0, \pm1, \pm2\}$ and $\Theta$ the local sidereal time (see Section~4 of \citealt{2010CQGra..27s4016A} for more information).

The amplitude $H_0$ is related to the standard GW amplitude $h_0$~\citep{1998PhRvD..58f3001J} as
\begin{equation}
\label{eq:h0_vs_H0}
H_0 = h_0 \sqrt{ \frac{1 + 6 \cos^2 \iota + \cos^4 \iota}{4} }\,,
\end{equation}
with $\iota$ being the angle between the rotation axis and line of sight. 
The polarization coefficients are given by
\begin{equation}
H_+ = \frac{ \cos(2\psi) - i \eta \sin(2\psi) }{ \sqrt{1 + \eta^2} },\,
H_\times = \frac{ \sin(2\psi) + i \eta \cos(2\psi) }{ \sqrt{1 + \eta^2} },
\end{equation}
being $\psi$ the CW polarization angle, and
\begin{equation}
\eta = -\frac{2 \cos \iota}{1 + \cos^2 \iota}\,.
\end{equation}
The matched filters $\hat{H}_{+/\times}$ are computed in the frequency domain as
\begin{equation}
\hat{H}_{+/\times} = \frac{ \mathbf{X} \cdot \mathbf{A}^{+/\times} }{ |\mathbf{A}^{+/\times}|^2 },
\label{eq:matched-filter}
\end{equation}
where $\mathbf{X}$ is the data five-vector
\begin{equation}
\mathbf{X} = \int_{T_{\text{obs}}} x(t) \, e^{-i \mathbf{k} \Theta} \, e^{-i \omega_{\rm gw} t} dt.
\end{equation}
Finally, the detection statistic is
\begin{equation}\label{eq:classicS}
S = |\mathbf{A}^+|^4 |\hat{H}_+|^2 + |\mathbf{A}^\times|^4 |\hat{H}_\times|^2\,,
\end{equation}
as defined in~\citet{Astone:2014dea}.
Note that recent works on the five-vector formalism~\citep{D’Onofrio_2025, Prix:2024xpl} have described the equivalence with the usual $\mathcal{F}$-statistic.

In this work, we focus on a narrowband approach where we search for CWs allowing for a small mismatch between the EM emission described by Equation~\ref{eq:timing-model} and the gravitational one.
We use the 5n-vector narrowband pipeline, widely used in several CW searches~\citep[e.g.][]{LIGOScientific:2021quq, 2017PhRvD..96l2006A, O2Narrowband, 2015PhRvD..91b2004A,LIGOScientific:2025kei, Mirasola:2025car}.
We explore a narrow frequency and spin-down range around twice the best-fit values from the pulsar ephemeris, to allow for a small mismatch between the EM and CW emissions, namely,
\begin{equation}
    f\in 2f_{\rm rot} [1-\delta,\, 1+\delta]
    \label{eq:range_freq_explored}
\end{equation}
generally setting $\delta=10^{-3}$~\citep{LIGOScientific:2021quq}, and a similar equation for $\dot{f}$ using the same value.
As a novelty of this work, we explore a corresponding additional range of values in the $\ddot{f}$ space.
During the implementation of this feature, we identified a bug in the spin-down correction of this pipeline used in~\citet{LIGOScientific:2025kei}.
After rerunning the analyses in~\citet{LIGOScientific:2025kei}, we find no additional candidates, and upper limits (ULs) are unchanged. 
Moreover, the results of~\citet{LIGOScientific:2025kei} are now superseded by those reported in  Section~\ref{sec:results}.

The method makes use of the short Fourier data base~\citep[SFDB;][]{Astone:2005fj}, which is a collection of short-duration (here we use 2048~s ones) fast Fourier transforms overlapped by half.
For each pulsar, we then extract a narrow frequency band from the SFDBs around the region of interest that is then inverse-Fourier-transformed to the time domain.

For every target, data are Doppler-corrected in the time domain using a nonuniform resampling method that is independent of the CW frequency, and then they are subsampled at a rate of 1~Hz.
For the first time, the pipeline can now target sources in binary systems, accounting for orbital modulation, following~\citet{Singhal_2019} and \citet{Amicucci:2025lqh}.

At this point, the time series are Fourier-transformed and matched-filtered (using Equation~\ref{eq:matched-filter}) to estimate the two CW polarizations using a template bank in the $f-\dot{f}-\ddot{f}$ space.
The template grid consists of all the resolved parameter-space points within the region of interest, spaced by the corresponding resolution calculated as~\citep{Astone:2014dea}
\begin{equation}
    \delta f = \frac{1}{\Tobs}\,, \qquad \delta \dot{f} = \frac{1}{\Tobs^2}\,, \qquad \delta \ddot{f} = \frac{2}{\Tobs^3}\,,
    \label{eq:spacing_grid_nb}
\end{equation}
with $\Tobs$ the data time span.
Note that these resolutions neglect correlations among parameters, resulting in more templates than needed~\citep{LIGOScientific:2008hqb}, though the overall impact on the computing cost is not significant.
Higher-order spin-down terms, if provided in the ephemerides, are fixed at twice the best-fit values to track the GW frequency evolution over time, without exploring any additional template~\citep{Astone:2014dea}.
When we consider a glitching pulsar, we split the data into two segments that exclude from the analysis the period around [ $t_{\rm g}$--1~day, $t_{\rm g}$+2~day ], with $t_{\rm g}$ the glitch epoch.
The segments are then analyzed independently.
If the pulsar glitched more than once, this process is repeated until we cover the entire observing time.

The matched filter results from different detectors are coherently combined following~\citet{Mastrogiovanni:2017xjr} to evaluate the detection statistic in Equation~\ref{eq:classicS}.
Then, marginalization over the spin-downs is performed by sorting the detection statistic values along the frequency axis and selecting the maximum in every $10^{-4}$~Hz band.
All templates with a detection statistic above a certain threshold, set by fixing the false-alarm probability ($\pfa$) at 1\% after taking the number of explored grid points into account, are followed up on\footnote{Note that our trial factors do not account for the number of studied targets.}. The noise-only distribution, used to set the threshold, is inferred with an exponential fit from the tail of the histogram of all statistic values which have not been selected as local maxima~\citep{Singhal_2019}. 

If no CW-related outlier is found, we calculate the 95\% confidence level (CL) ULs $h_0^{95\%}$ by injecting simulated signals in real data following the procedure detailed in Appendix~A of~\citet{2017PhRvD..96l2006A}, with the only difference being that we simulate signals with random $\ddot f$ within the explored range.
A summary of the procedure is given here in Appendix~\ref{app:hul}.

\section{Search details and results\label{sec:results}}
In this section, we present and discuss our results using LIGO O4ab data.
We do not report a detection.
The search identified a set of outliers for PSRs J0117+5914 and J1826\textminus1334, for which we reject an astrophysical origin, as justified below.
We first describe the target selection procedure and search setup in Section~\ref{sec:targets}.
Then, we detail our results in Section~\ref{sec:upper_limits} which are then interpreted in terms of astrophysical constraints in Section~\ref{astrophys_intr}.

\subsection{Target selection and search setup\label{sec:targets}}

The O4ab narrowband search selected all the targets from the set presented in Section~\ref{sec:em_data}.
We focus on those pulsars with $h_0^{\text{sd}}$ above or within a factor of 4 below the expected sensitivity calculated as~\citep{Astone:2014dea,D’Onofrio_2025}
\begin{equation}
    h_0^{\rm sens} \approx \mathcal{C}\, \sqrt{ \left(\sum\limits_{i=1}^n \frac{\mu^i\,\Tobs^i}{S_i}\right)^{-1} }\,,
    \label{eq:hsens}
\end{equation}
where $\mu^i$ and $S_i$ are, respectively, the duty cycle and the harmonic average power spectral density (PSD) for the $i$th detector.
The prefactor $\mathcal{C}$ in Equation~\ref {eq:hsens} mildly depends on the number of explored templates $\Ntrial$ for each target~\citep{Astone:2014dea,D’Onofrio_2025}.
In our plots, we consider $\mathcal{C}\in[20, 26.5]$ which roughly corresponds to $\Ntrial\in[10^3, \,10^9]$.

\begin{figure}[!ht]
    \centering
    \includegraphics[width=\columnwidth]{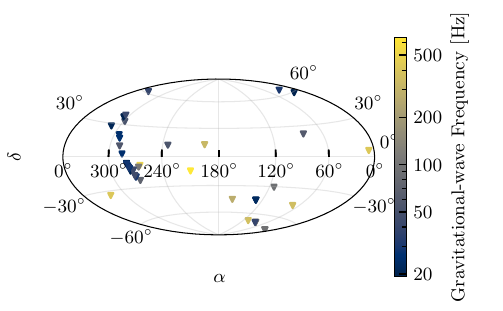}
    \caption{
    Sky location in equatorial coordinates of the selected targets. The color scale indicates the target's $f_{\rm gw}$.}
    \label{fig:sky_pos_targets}
\end{figure}

Of the 39 targets within this range, 7 isolated targets glitched within the O4ab period\footnote{Note that Table~\ref{tab:glitches_O4ab} reports nine glitching pulsars, but two glitched \textit{before} the beginning of the run, therefore we do not have to perform separate pre- and post-glitch analyses.} (see Section~\ref{sec:glitches}), and 10 are in a binary system.
Figure~\ref{fig:sky_pos_targets} shows 
the sky location of the pulsars targeted in this work.

For PSR J0540\textminus6919, we analyzed data only after the second timing event, since both are very close to the beginning of the run~\citep{Espinoza:2024cwk}.

Following~\citet{Mirasola:2024kll}, we discard one pulsar, PSR J1231\textminus1411, in a binary system for which the ephemeris does not provide all the Keplerian parameters needed to correct for the source's orbital motion.
For each of the other targets in binary systems, we calculate the longest considerable time span ($\Tobs^{\rm max}$) within which a single fully coherent search can be applied without needing to account for uncertainties on orbital parameters~\citep{Mirasola:2024kll}.
For four of the pulsars, PSRs J0437\textminus4715, J1045\textminus4509, J1737\textminus0811, and J1745\textminus0952, this calculation returns segments much shorter ($1~\mathrm{day}\lesssim\Tobs^{\rm max}\lesssim90~\mathrm{days}$) than $\Tobs^{\rm O4ab}\approx 610$~days.
We exclude these four targets since the 5n-vector narrowband pipeline does not cover orbital uncertainties.
On the other hand, we have split $\Tobs$ into three separate segments for pulsar J1400\textminus1431, since $\Tobs^{\rm max}\sim 200$~days. We have not considered the second segment, as it includes the commissioning break, and the $\mu\,\Tobs$ would not be enough to perform the analysis.
For the other pulsars in a binary system, we have analyzed the entire dataset coherently.\\
Due to storage limitations, we had to restrict the analysis to $\delta = 10^{-4}$ for  PSRs J0534+2200 (Crab pulsar), J0540\textminus6919, J0205+6449, and J2229+6114.

As a result, we present in the following section the search for CWs emitted by 34 selected targets. 

\subsection{Outliers and ULs\label{sec:upper_limits}}

\begin{figure}[!ht]
    \centering
    \includegraphics[width=\columnwidth]{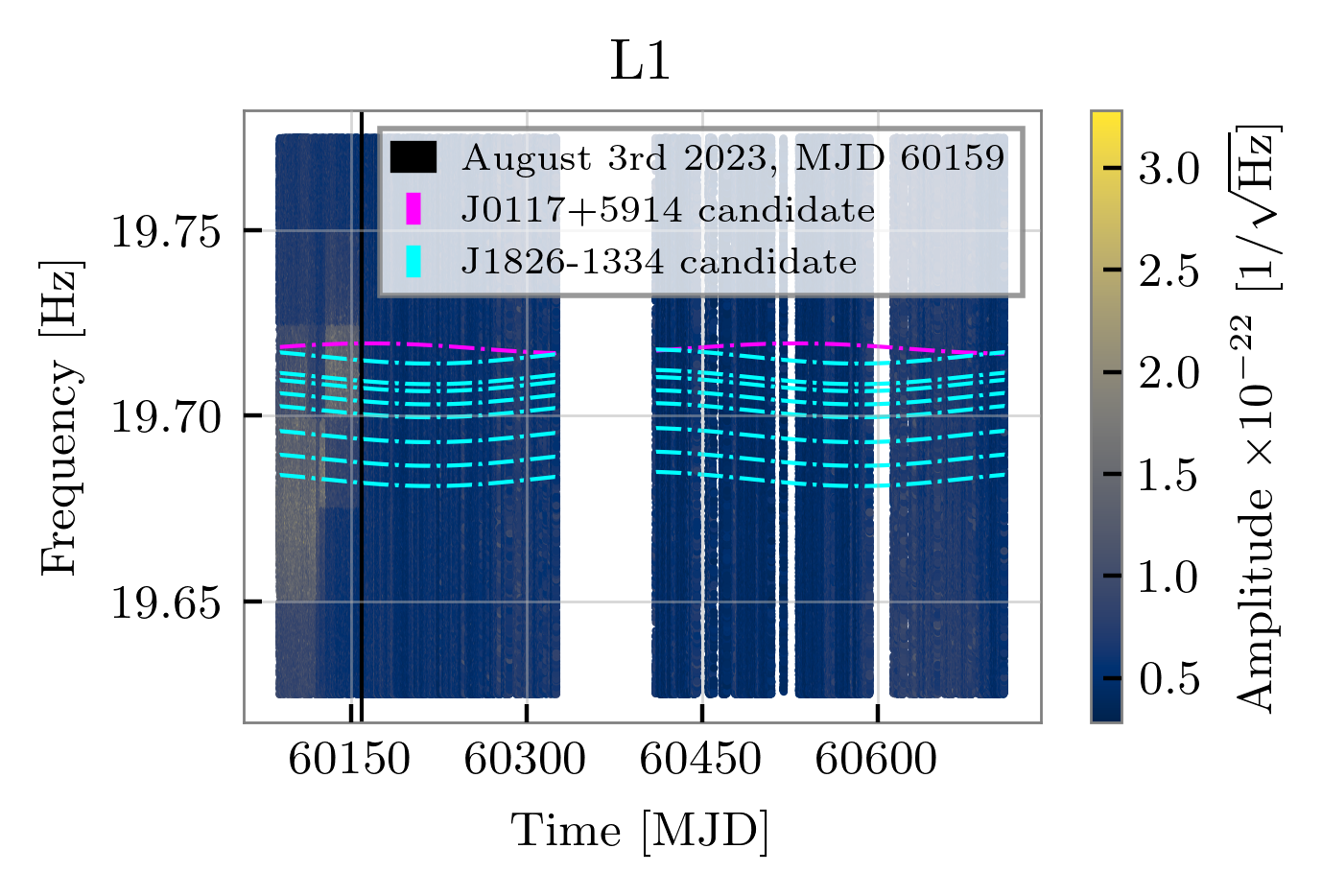}
    \caption{L1 spectrogram computed every 2048~s over the O4ab run. The PSRs J0117+5914 and J1826\textminus1334 outlier tracks, calculated with Equation~\ref{eq:freq_track}, are displayed in fuchsia and cyan, respectively. During the first part of the dataset, the spectrum is affected by a known noise transient reported in~\citet{transient_noise_L1} whose strength reduces after the MJD 60159 (vertical black line), see text for more details. 
    }
    \label{fig:spectrogram_outliers}
\end{figure}

Only for two targets we report outliers (i.e., inferred \textit{p}-value $<1\%$ after accounting for $\Ntrial$): one for PSR J0117+5914 and eight for PSR J1826\textminus1334.
The two pulsars have a similar rotational frequency ($\sim 19.7$~Hz) that does not overlap with known line-like features reported in~\citet{o4ab_linelist}, though L1 data are affected by a known transient noise reported~\citep{transient_noise_L1}, which happens to cross the outliers' band.
The noise is related to a $\sim$3.4~Hz line, which interacts with the $\sim$16.3~Hz calibration line. 
The interaction lost strength on 2023 August 3 (MJD 60159).
This transient behavior, reported in~\citet{transient_noise_L1}, is readily shown in 
Figure~\ref{fig:spectrogram_outliers}, where we display the spectrogram for the L1 detector along the run built with segments of 2048~s.
For comparison, we superimpose the outliers' frequency tracks encompassing Doppler and spin-down effects as~\citep{Krishnan:2004sv}
\begin{equation}
    f(t) = \left[f_0 + \dot{f}_0 \, (t-t_0) + \frac{1}{2} \ddot{f}_0 \, (t-t_0)^2 \right] \left(1+\frac{\vec{v}_{\rm det}\cdot \vec{n}}{c}\right),
    \label{eq:freq_track}
\end{equation}
with the parameters $f_0,\,\dot{f}_0,\,\ddot{f}_0$ as those of the outliers, $t$ the time at the detector, $\Vec{v}_{\rm det}$ the detector's velocity in the solar system barycentre, and $\Vec{n}$ the source sky location.

\startlongtable
\centerwidetable
\begin{deluxetable*}{lcccccccccc}

\setlength\tabcolsep{3pt}
\tablecaption{Results table obtained with the method described in Section~\ref{sec:signal_and_method}.\label{tab:narrowband_results}}

\tabletypesize{\scriptsize}

\tablehead{
\colhead{Pulsar Name} & 
\colhead{$f_{\rm rot}$} & 
\colhead{$\dot{f}_{\rm rot}$} &
\colhead{$\ddot{f}_{\rm rot}$} &
\colhead{$d$} & 
\colhead{$h_0^{95\%}$} &
\colhead{$h_0^{95\%}/h_0^{sd}$} & 
\colhead{$\mathcal{D}$} & 
\colhead{$\varepsilon^{95\%}$} & 
\colhead{N$_{\mathrm{trials}}$} & 
\colhead{Distance Ref.} \\
\colhead{J2000} & 
\colhead{Hz} & 
\colhead{Hz/s} & 
\colhead{Hz/s$^2$} & 
\colhead{kpc} & 
\colhead{$\times 10^{-26}$} & 
\colhead{~} & 
\colhead{~} & 
\colhead{~} & 
\colhead{$\times 10^6$} & 
\colhead{~}}

\startdata
J0030+0451\tablenotemark{$\rho$} & 205.53 & $-4.3 \times 10^{-16}$ & - & 0.32 & 1.52 & 4.21 & 233.8 & $2.7 \times 10^{-8}$ & 46 & \citet{Lommen:2000yt}\\
J0058\textminus7218\tablenotemark{$\beta$} & 45.94 & $-6.2 \times 10^{-11}$ & $6.6 \times 10^{-21}$ & 59.70 & 3.16 & 2.01 & 154.2 & $2.1 \times 10^{-4}$ & 701 & \citet{2021MNRAS.507L...1M}\\
J0058\textminus7218 pg\tablenotemark{$\beta$} & 45.94 & $-6.2 \times 10^{-11}$ & $4.9 \times 10^{-21}$ & 59.70 & 3.21 & 2.04 & 152.0 & $2.1 \times 10^{-4}$ & 6047 & \citet{2021MNRAS.507L...1M}\\
J0117+5914\tablenotemark{$\gamma$} & 9.86 & $-5.7 \times 10^{-13}$ & $-8.4 \times 10^{-24}$ & 1.77 & 26.69 & 2.44 & 267.5 & $1.1 \times 10^{-3}$ & 55 & \citet{Yao+17}\\
J0205+6449$^*$\tablenotemark{$\gamma$} & 15.19 & $-4.5 \times 10^{-11}$ & $1.7 \times 10^{-20}$ & 3.20 & 6.67 & 0.15 & 244.0 & $2.2 \times 10^{-4}$ & 50 & \citet{2002ApJ...568..226M}\\
J0534+2200$^*$\tablenotemark{$\gamma$} & 29.57 & $-3.7 \times 10^{-10}$ & $1.8 \times 10^{-20}$ & 2.00 & 3.05 & 0.02 & 232.0 & $1.6 \times 10^{-5}$ & 753 & \citet{2012ApSS.341...43W}\\
J0537\textminus6910\tablenotemark{$\beta$} & 61.89 & $-2.0 \times 10^{-10}$ & $5.8 \times 10^{-21}$ & 49.70 & 3.63 & 1.25 & 121.5 & $1.1 \times 10^{-4}$ & 896 & \citet{1998ApJ...499L.179M}\\
J0537\textminus6910 pg1\tablenotemark{$\beta$} & 61.89 & $-2.0 \times 10^{-10}$ & $5.8 \times 10^{-21}$ & 49.70 & 4.13 & 1.42 & 106.8 & $1.3 \times 10^{-4}$ & 1447 & \citet{1998ApJ...499L.179M}\\
J0537\textminus6910 pg2\tablenotemark{$\beta$} & 61.89 & $-2.0 \times 10^{-10}$ & $1.4 \times 10^{-20}$ & 49.70 & 3.85 & 1.32 & 114.7 & $1.2 \times 10^{-4}$ & 4057 & \citet{1998ApJ...499L.179M}\\
J0537\textminus6910 pg3\tablenotemark{$\beta$} & 61.89 & $-2.0 \times 10^{-10}$ & $9.0 \times 10^{-21}$ & 49.70 & 4.41 & 1.51 & 100.1 & $1.4 \times 10^{-4}$ & 252 & \citet{1998ApJ...499L.179M}\\
J0540\textminus6919$^*$\tablenotemark{$\beta$} & 19.64 & $-2.5 \times 10^{-10}$ & $4.7 \times 10^{-21}$ & 49.70 & 4.99 & 0.86 & 205.5 & $1.5 \times 10^{-3}$ & 226 & \citet{10.1093/mnras/stab1501}\\
J0711\textminus6830\tablenotemark{$\nu$} & 182.12 & $-4.9 \times 10^{-16}$ & - & 0.11 & 1.43 & 1.15 & 236.5 & $1.1 \times 10^{-8}$ & 83 & \citet{Yao+17}\\
J0737\textminus3039A$^\dagger$\tablenotemark{$\gamma$} & 44.05 & $-3.4 \times 10^{-15}$ & - & 1.10 & 2.00 & 3.09 & 250.2 & $2.7 \times 10^{-6}$ & 27 & \citet{Burgay:2003jj}\\
J0835\textminus4510\tablenotemark{$\delta$} & 11.18 & $-1.6 \times 10^{-11}$ & $2.0 \times 10^{-22}$ & 0.28 & 23.44 & 0.07 & 170.7 & $1.2 \times 10^{-4}$ & 213 & \citet{2012ApJ...755...39V}\\
J0835\textminus4510 pg\tablenotemark{$\delta$} & 11.18 & $-1.6 \times 10^{-11}$ & $2.0 \times 10^{-22}$ & 0.28 & 24.58 & 0.07 & 162.8 & $1.3 \times 10^{-4}$ & 107 & \citet{2012ApJ...755...39V}\\
J0835\textminus4510 pgMt\tablenotemark{$\epsilon$} & 11.18 & $-1.6 \times 10^{-11}$ & $-2.3 \times 10^{-21}$ & 0.28 & 39.02 & 0.11 & 102.6 & $2.1 \times 10^{-4}$ & 11 & \citet{2012ApJ...755...39V}\\
J1300+1240$^\dagger$\tablenotemark{$\alpha$} & 160.81 & $-3.0 \times 10^{-15}$ & - & 0.71 & 1.98 & 4.06 & 176.2 & $1.3 \times 10^{-7}$ & 101 & \citet{Wolszczan_2000}\\
J1400\textminus1431\_1$^\dagger$\tablenotemark{$\alpha$} & 324.23 & $-7.6 \times 10^{-16}$ & - & 0.28 & 2.68 & 6.04 & 148.1 & $1.7 \times 10^{-8}$ & 48 & \citet{2013ApJ...768...85R}\\
J1400\textminus1431\_2$^\dagger$\tablenotemark{$\alpha$} & 324.23 & $-7.6 \times 10^{-16}$ & - & 0.28 & 3.10 & 6.96 & 128.4 & $1.9 \times 10^{-8}$ & 48 & \citet{2013ApJ...768...85R}\\
J1537+1155$^\dagger$\tablenotemark{$\gamma$} & 26.38 & $-1.7 \times 10^{-15}$ & $6.2 \times 10^{-29}$ & 0.93 & 3.15 & 4.57 & 231.3 & $1.0 \times 10^{-5}$ & 49 & \citet{Fonseca_2014}\\
J1756-2251$^\dagger$\tablenotemark{$\gamma$} & 35.14 & $-1.3 \times 10^{-15}$ & $3.0 \times 10^{-27}$ & 0.73 & 2.39 & 3.63 & 235.0 & $3.3 \times 10^{-6}$ & 22 & \citet{2004MNRAS.355..147F}\\
J1809\textminus1917\tablenotemark{$\gamma$} & 12.08 & $-3.7 \times 10^{-12}$ & $6.2 \times 10^{-23}$ & 3.27 & 12.91 & 0.94 & 240.4 & $6.8 \times 10^{-4}$ & 326 & \citet{Yao+17}\\
J1811\textminus1925\tablenotemark{$\beta$} & 15.45 & $-1.1 \times 10^{-11}$ & $1.8 \times 10^{-22}$ & 5.00 & 6.75 & 0.50 & 227.9 & $3.3 \times 10^{-4}$ & 1175 & \citet{1997ApJ...489L.145T}\\
J1813\textminus1246\tablenotemark{$\zeta$} & 20.80 & $-7.6 \times 10^{-12}$ & $2.6 \times 10^{-22}$ & 2.63 & 7.49 & 0.40 & 124.2 & $1.1 \times 10^{-4}$ & 472 & \citet{2009Sci...325..840A}\\
J1813\textminus1246 pg\tablenotemark{$\zeta$} & 20.80 & $-7.6 \times 10^{-12}$ & $2.6 \times 10^{-22}$ & 2.63 & 5.61 & 0.30 & 165.7 & $8.1 \times 10^{-5}$ & 472 & \citet{2009Sci...325..840A}\\
J1813\textminus1749\tablenotemark{$\eta$} & 22.34 & $-6.3 \times 10^{-11}$ & - & 6.15 & 3.76 & 0.17 & 224.3 & $1.1 \times 10^{-4}$ & 3312 & \citet{Yao+17}\\
J1826\textminus1334\tablenotemark{$\gamma$} & 9.85 & $-7.3 \times 10^{-12}$ & $2.1 \times 10^{-22}$ & 3.61 & 27.22 & 1.42 & 263.4 & $2.4 \times 10^{-3}$ & 513 & \citet{Yao+17}\\
J1828\textminus1101\tablenotemark{$\gamma$} & 13.88 & $-2.8 \times 10^{-12}$ & $-3.9 \times 10^{-23}$ & 4.77 & 9.58 & 1.25 & 299.6 & $5.6 \times 10^{-4}$ & 95 & \citet{Yao+17}\\
J1831\textminus0952\tablenotemark{$\gamma$} & 14.86 & $-1.8 \times 10^{-12}$ & $7.9 \times 10^{-24}$ & 3.68 & 7.04 & 0.92 & 256.6 & $2.8 \times 10^{-4}$ & 71 & \citet{Yao+17}\\
J1833\textminus0827\tablenotemark{$\gamma$} & 11.72 & $-1.3 \times 10^{-12}$ & $-1.7 \times 10^{-24}$ & 4.50 & 13.75 & 2.34 & 240.9 & $1.1 \times 10^{-3}$ & 125 & \citet{10.1111/j.1365-2966.2011.19022.x}\\
J1837\textminus0604\tablenotemark{$\gamma$} & 10.38 & $-4.9 \times 10^{-12}$ & $3.5 \times 10^{-22}$ & 4.78 & 20.16 & 1.74 & 264.2 & $2.1 \times 10^{-3}$ & 123 & \citet{Yao+17}\\
J1838\textminus0655\tablenotemark{$\beta$} & 14.18 & $-1.0 \times 10^{-11}$ & $3.0 \times 10^{-22}$ & 6.60 & 8.48 & 0.83 & 248.1 & $6.6 \times 10^{-4}$ & 335 & \citet{Gotthelf:2008yt}\\
J1856+0245\tablenotemark{$\gamma$} & 12.36 & $-9.5 \times 10^{-12}$ & $1.8 \times 10^{-22}$ & 6.32 & 12.63 & 1.13 & 242.3 & $1.2 \times 10^{-3}$ & 276 & \citet{Yao+17}\\
J1913+1011\tablenotemark{$\gamma$} & 27.85 & $-2.6 \times 10^{-12}$ & $-1.7 \times 10^{-23}$ & 4.61 & 2.77 & 0.51 & 241.1 & $3.9 \times 10^{-5}$ & 180 & \citet{Yao+17}\\
J1925+1720\tablenotemark{$\gamma$} & 13.22 & $-1.8 \times 10^{-12}$ & $1.8 \times 10^{-23}$ & 5.06 & 9.53 & 1.61 & 252.4 & $6.5 \times 10^{-4}$ & 63 & \citet{Yao+17}\\
J1935+2025\tablenotemark{$\gamma$} & 12.48 & $-9.5 \times 10^{-12}$ & $2.9 \times 10^{-22}$ & 4.59 & 12.22 & 0.80 & 236.9 & $8.5 \times 10^{-4}$ & 278 & \citet{Yao+17}\\
J1952+3252\tablenotemark{$\gamma$} & 25.29 & $-3.7 \times 10^{-12}$ & $9.2 \times 10^{-22}$ & 3.00 & 3.16 & 0.31 & 228.8 & $3.5 \times 10^{-5}$ & 684 & \citet{Zeiger_2008}\\
J2021+3651\tablenotemark{$\gamma$} & 9.64 & $-8.9 \times 10^{-12}$ & $-2.5 \times 10^{-22}$ & 1.80 & 40.04 & 0.93 & 209.8 & $1.8 \times 10^{-3}$ & 25 & \citet{2002ApJ...577L..19R}\\
J2021+3651 pg\tablenotemark{$\gamma$} & 9.64 & $-8.9 \times 10^{-12}$ & $-2.5 \times 10^{-22}$ & 1.80 & 42.24 & 0.98 & 199.0 & $1.9 \times 10^{-3}$ & 190 & \citet{2002ApJ...577L..19R}\\
J2022+3842\tablenotemark{$\beta$} & 20.57 & $-3.7 \times 10^{-11}$ & $6.5 \times 10^{-21}$ & 10.00 & 5.44 & 0.51 & 175.5 & $3.0 \times 10^{-4}$ & 1757 & \citet{2011ApJ...739...39A}\\
J2043+2740\tablenotemark{$\gamma$} & 10.40 & $-1.3 \times 10^{-13}$ & $3.1 \times 10^{-23}$ & 1.48 & 19.67 & 3.19 & 268.8 & $6.4 \times 10^{-4}$ & 6 & \citet{Yao+17}\\
J2124\textminus3358\tablenotemark{$\nu$} & 202.79 & $-8.5 \times 10^{-16}$ & - & 0.41 & 1.53 & 3.80 & 225.1 & $3.6 \times 10^{-8}$ & 127 & \citet{Reardon+16}\\
J2229+6114$^*$\tablenotemark{$\gamma$} & 19.35 & $-2.9 \times 10^{-11}$ & $2.7 \times 10^{-21}$ & 3.00 & 4.24 & 0.13 & 241.6 & $8.0 \times 10^{-5}$ & 41 & \citet{2001ApJ...552L.125H}\\
 \hline
\enddata
\tablecomments{The nomenclature ``pg'' identifies the post-glitch analyses, ``pgMt'' identifies the second post-glitch Vela analysis using the Mount Pleasant ephemeris; more information is given in the text. Starred pulsars have been analysed using $\delta = 10^{-4}$. The marker $^\dagger$ identifies pulsars in a binary system. Frequencies and spin-down parameters are referred to the starting day of the run. The following is a list of references for pulsar ephemeris data used in this analysis: Nancay: $\alpha$, NICER: $\beta$, JBO: $\gamma$, IAR: $\delta$, MeerKAT: $\nu$, CHIME: $\rho$, Fermi-LAT: $\zeta$, Chandra: $\eta$, Mount Pleasant: $\epsilon$.}
\tablerefs{The last column lists references for pulsar distances, and they should be consulted for information on their associated uncertainty.}

\end{deluxetable*}

\begin{figure*}[!ht]
    \centering
    \includegraphics[width=0.85\textwidth]{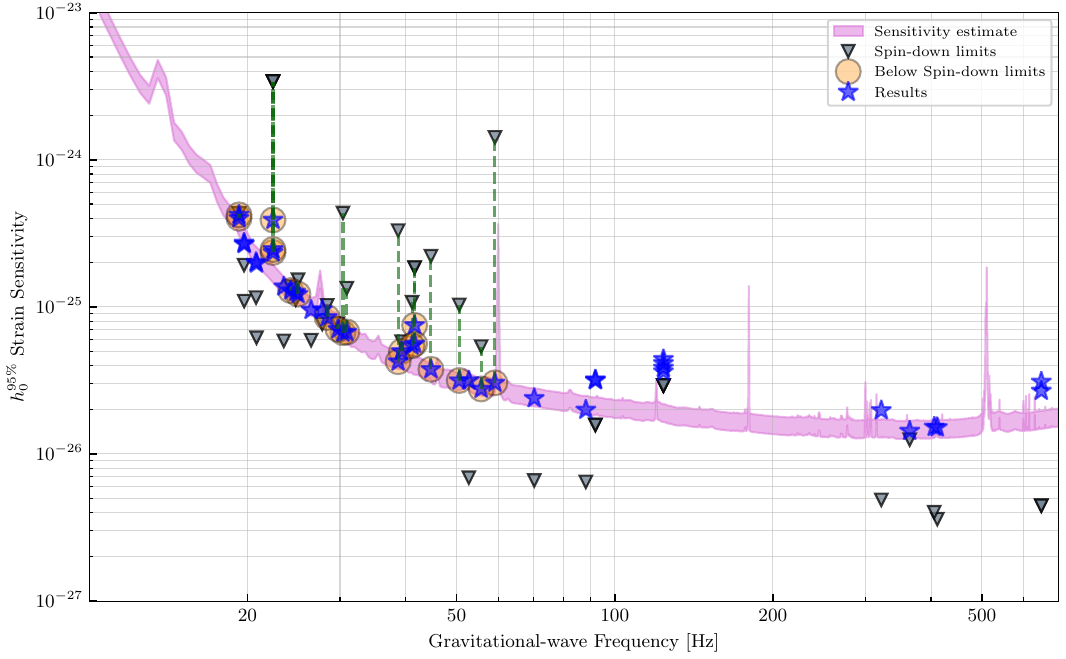}
    \caption{Expected sensitivity of the narrowband search using the O4ab (shaded pink region) dataset from the two LIGO detectors. The curve is compared with the spin-down limits (triangles) and the 95\% CL ULs (stars) averaged over all the $10^{-4}$~Hz bands for each source. ULs below the spin-down limit are highlighted with orange circles. Superimposed blue stars correspond to the multiple analyses of glitching pulsars.}
    \label{fig:NB_ULs}
\end{figure*}

Since all the tracks are crossing the polluted band, their astrophysical origin is discarded.

As a result, we have calculated ULs at the 95\% CL for each of the analyzed pulsars.
In Table~\ref{tab:narrowband_results} we report the per-pulsar ULs, and in Figure~\ref{fig:NB_ULs} we compare $h_0^{95\%}$ with the expected sensitivity estimated through Equation~\ref{eq:hsens}.

For PSR J0835\textminus4510 (the Vela pulsar), one of our most interesting targets, we performed an independent post-glitch cross-check analysis with a Mount Pleasant ephemeris covering $\Tobs\sim117$~days of O4b.  
After correcting for the shorter $\Tobs$ compared with $\Tobs \sim 270$~days for the IAR ephemeris and the corresponding number of templates, the two results are in agreement. See Table~\ref{tab:narrowband_results} for numerical values.
Constraints on GW emission from this target using O4 data have also been presented in~\citet{KAGRA:2025qeh}, who searched for post-glitch transient signals.
Results for signals lasting the full 120~days considered in that paper are in broad agreement with the Mount Pleasant ephemeris UL reported here.
However, the two analyses considered different time evolution of the signal and different template banks, so there is no exact correspondence. 

We also compare $h_0^{95\%}$ with the spin-down limit $h_0^{\rm sd}$ of each source using a moment of inertia $I_{\rm zz}$~=~$10^{38}$~kg~m$^2$ in Equation~\ref{eq:spin-down}.
In this work, we have surpassed the spin-down limit in 20 analyses (counting pre- and post-glitch searches as separate analyses), with our most stringent constraint being a ratio of $\sim2\%$ between our UL and Crab's $h_{0}^{\rm sd}$.

\begin{figure}[!ht]
    \centering
    \includegraphics[width=\columnwidth]{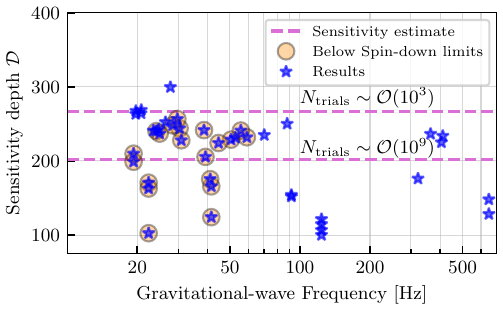}
    \caption{Sensitivity depths calculated with Equation~\ref{eq:sens_depth} using the results in Figure~\ref{fig:NB_ULs}. The two horizontal dotted lines highlight the theoretical sensitivity for the reported number of explored templates, calculated with Equations~\ref{eq:hsens} and~\ref{eq:sens_depth}.}
    \label{fig:nb_ul_depth}
\end{figure}

From the ULs in Figure~\ref{fig:NB_ULs}, we calculate the dimensionless sensitivity depth $\mathcal{D}$~\citep{Behnke:2014tma,Dreissigacker:2018afk,WETTE2023102880} as
\begin{equation}
    \mathcal{D}=\frac{\sqrt{S_h\cdot{\rm Hz}}}{h_0^{95\%}}
    \label{eq:sens_depth}
\end{equation}
where $S_h$ is the power spectral density, harmonic-averaged over time and detectors. See~\citet{Behnke:2014tma} and \citet{Dreissigacker:2018afk} for additional references on the sensitivity depth.
In this work, we reached a median depth of $\mathcal{D}\approx 230$ (not including glitching pulsar analyses), in agreement with previous searches~\citep{Wette:2020air}.
Note that the distribution of $\mathcal{D}$ in Figure~\ref{fig:nb_ul_depth}, where we report the per-pulsar depths, can be tracked back to either $\Ntrial$ or a glitching pulsar analysis (i.e., where we split the dataset into shorter segments).

\begin{figure}[ht]
    \centering
    \includegraphics[width=0.47\textwidth]{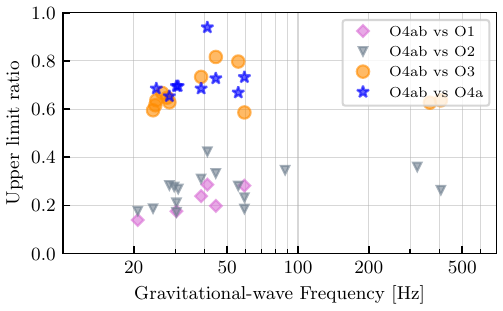}
    \caption{ULs in Figure~\ref{fig:NB_ULs} of nonglitching pulsars divided by our previous 5n-vector narrowband ULs. In particular, blue stars compare with the O4a-only analysis of~\citet{LIGOScientific:2025kei}, orange circles with O3~\citep{LIGOScientific:2021quq}, gray triangles with O2~\citep{O2Narrowband}, and O1 purple squares from~\citet{2017PhRvD..96l2006A}. Note that multiple analyses might have targeted the same pulsar; we show all the comparisons to highlight the significant improvements from one observing run to the other.}
    \label{fig:comparison_w_prev_runs}
\end{figure}

In this work, we improve our previous constraints on the CW emission for all the considered nonglitching targets, as shown in Figure~\ref{fig:comparison_w_prev_runs}.
If we compare with O4a~\citep{LIGOScientific:2025kei} and O3~\citep{LIGOScientific:2021quq} analyses, which have a comparable sensitivity~\citep{LIGOScientific:2025kei}, we see an improvement of about $\sim1/\sqrt{2}$ related to the doubled $\Tobs$ (see Equation~\ref{eq:hsens}).
If we compare with O2~\citep{O2Narrowband} and O1~\citep{2017PhRvD..96l2006A}, the improvement is even larger thanks to the detectors' upgrades and a longer run.

\subsection{Astrophysical interpretation\label{astrophys_intr}}

\begin{figure*}[ht]
    \centering
    \includegraphics[width=0.47\textwidth]{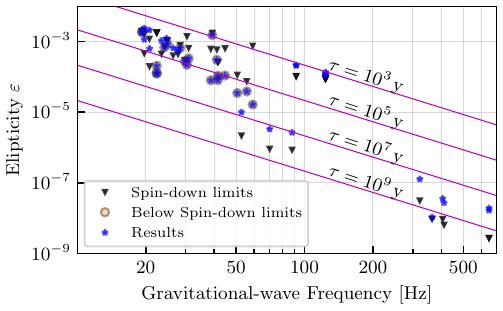}
    \includegraphics[width=0.47\textwidth]{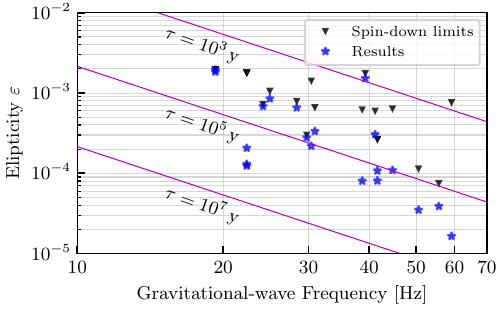}
    \caption{ULs on the star's ellipticity $\varepsilon$ compared with the spin-down limits. While the left panel shows our results for all the pulsars, the right panel focuses only on those for which the observational UL is stricter than the spin-down limit (i.e., physically constraining the fraction of rotational energy emitted through the CW channel). The values are compared with contour lines of equal characteristic age $\tau = f / 4 |\dot{f}|$, assuming that GW emission alone is causing the spin-down.}
    \label{fig:ellipticity_upper_limits}
\end{figure*}

In Figure~\ref{fig:ellipticity_upper_limits} we recast our results as ULs on ellipticities, as defined in Equation~\ref{eq:epsilon_definition}; these are also reported in Table~\ref{tab:narrowband_results}.  In the figure, we also show lines of constant characteristic age $\tau = f / 4 |\dot f|$, as would be relevant if spin-down were driven entirely by GW emission from a constant mass quadrupole.   This plot provides a natural place to interpret the astrophysical significance of our results.  The ULs naturally divide into two subclasses.  

For the younger pulsars, with $f \lesssim 100$ Hz, the ellipticity limits lie mainly in the range $10^{-3} \lesssim \varepsilon \lesssim 10^{-5}$, with a few ULs of a few times $10^{-6}$.  All of the pulsars for which we have beaten the energy-based spin-down limit belong to this class.  The limit is surpassed by the greatest margin for the Crab pulsar (J0534+2200), for which we find $h_0^{95\%} / h_0^{\rm sd} \approx 0.02$, corresponding to no more than $0.02^2 \approx 0.04\%$ of its spin-down energy budget being radiated in the GW channel.

For the older millisecond pulsars, with $f \gtrsim 100$ Hz, our ULs on $h_0$  lead to upper bounds on ellipticity lying roughly in the range $10^{-8} \lesssim \varepsilon \lesssim 10^{-7}$.  While smaller than the ellipticity constraints quoted above for the young pulsars, we do \emph{not} beat the spin-down ULs for any of the millisecond pulsars, and thus their ellipticity constraints are not physically informative.
The closest to the spin-down limit was  PSR J0711\textminus6830, for which we report $h_0^{95\%} / h_0^{\rm sd} \approx 1.15$.  Note, however, our calculations of the spin-down limits assume a canonical moment of inertia $I_{zz} = 10^{38}$ kg m$^2$.  For realistic NS masses and equations of state, values as large as $I_{zz} \approx 3 \times 10^{38}$ kg m$^2$ are possible \citep{2008ApJ...685..390W}.  In this case, the spin-down limits in $h_0^{\rm sd}$ are decreased by a factor $\sqrt 3$ (see Equation~\ref{eq:spin-down}), bringing this millisecond pulsar into the range of potential detectability, a significant achievement for a narrowband search.

To place our ULs in further context, theoretical estimates of maximum ellipticities supported by strains in NS crusts are around $10^{-6}$, somewhat below our best ULs for young pulsars; see e.g.\ \citet{Pitkin:2011sc, Johnson-McDaniel:2012wbj, 2000MNRAS.319..902U, gittins2021modelling, mh_22}. Theoretical estimates of the ellipticities produced by magnetic fields depend upon the poorly constrained geometry and strength of the \emph{internal} magnetic field $B_{\rm int}$, with an estimate of $\varepsilon \sim 10^{-8} (B_{\rm int}/ 10^{12}$ G) for a superconducting interior (see  \citet{Abbott_2020} for discussion and references).  For such magnetic mountains to reach the level of our ULs for the young pulsars, the internal field strength would have to be several orders of magnitude greater than the external field strength $\sim 10^{12}$ G typical of such pulsars \citep{lyne_gs_2012}. 

However, mountains supported by less conventional matter phases, e.g.\ solid quark phases or the colour-flavour-locked (CFL) phase, can be as large as $\sim 10^{-3}$; see e.g.,  \citet{2005PhRvL..95u1101O, gjs_12}.  While exotic, such stars are potentially particularly relevant to narrowband searches, where a possible emission mechanism is from a nonaxisymmetric solid core, whose rotation is not perfectly coupled to that of the crust, allowing for a GW emission offset slightly from (twice) the radio pulsation frequency \citep{2008ApJ...683L..45A}. By this measure, all of our ULs that are tighter than the spin-down limits are of astrophysical interest, but not directly comparable with those set by targeted searches (e.g., \citet{LIGOScientific:2025kei}) due to the intrinsic assumptions on the CW emission being mismatched (narrowband) or locked (targeted) to the EM emission.

\section{Conclusions\label{sec:conclusions}}
In this work, we present a search for CW signals from a set of 34 known pulsars using O4ab data from the two LIGO detectors, allowing for a mismatch between the EM and GW emissions.
This is the largest set of targets ever considered for a narrowband search.

We performed the CW search using the 5n-vector narrowband pipeline, based on matched filtering in the frequency domain.
Narrowband searches allow a small mismatch between the EM and CW emissions, unlike targeted searches~\citep[e.g.,][]{LIGOScientific:2025kei}. This implies that the results presented here are more robust in cases where the two signals are not phase-locked, e.g., for glitching pulsars, which are likely to have a superfluid interior flow.

For the first time, we study pulsars in binary systems in a narrowband search.
Thanks to other improvements of the 5n-vector pipeline, it can now explore an additional range of values in the $\ddot{f}$ space.

We do not report a detection as all statistically significant outliers were associated with instrumental disturbances and vetoed.
Therefore, we set ULs on the GW strain amplitude and ellipticity from each target.
For 20 analyses, we report an UL below the theoretical spin-down limit.
Overall, we improve our constraints on the CW emission from all the targets as shown in Figure~\ref{fig:comparison_w_prev_runs}.
Our tightest constraint is for the Crab pulsar, for which our UL on the CW amplitude is $\lesssim 2\%$ of its spin-down limit, corresponding to no more than $\sim0.04\%$ of its spin-down energy budget being radiated in the GW channel.

Data products associated with this work are collected in the LIGO-DCC document T2600052.

\section*{Acknowledgements}

%
\newif\ifcoreonly\coreonlyfalse
\newif\ifkagra\kagratrue
\newif\ifheader\headerfalse
\ifheader
\begin{center}{\bf\Large
\ifkagra
Conference proceedings acknowledgements for \\ the LIGO Scientific Collaboration, the Virgo Collaboration and the KAGRA Collaboration
\else
Conference proceedings acknowledgements for \\ the LIGO Scientific Collaboration and the Virgo Collaboration
\fi
}\end{center}
\fi
This material is based upon work supported by NSF's LIGO Laboratory, which is a
major facility fully funded by the National Science Foundation.
The authors also gratefully acknowledge the support of
the Science and Technology Facilities Council (STFC) of the
United Kingdom, the Max-Planck-Society (MPS), and the State of
Niedersachsen/Germany for support of the construction of Advanced LIGO 
and construction and operation of the GEO\,600 detector. 
Additional support for Advanced LIGO was provided by the Australian Research Council.
The authors gratefully acknowledge the Italian Istituto Nazionale di Fisica Nucleare (INFN),  
the French Centre National de la Recherche Scientifique (CNRS) and
the Netherlands Organization for Scientific Research (NWO)
for the construction and operation of the Virgo detector
and the creation and support  of the EGO consortium. 
The authors also gratefully acknowledge research support from these agencies as well as by 
the Council of Scientific and Industrial Research of India, 
the Department of Science and Technology, India,
the Science \& Engineering Research Board (SERB), India,
the Ministry of Human Resource Development, India,
the Spanish Agencia Estatal de Investigaci\'on (AEI),
the Spanish Ministerio de Ciencia, Innovaci\'on y Universidades,
the European Union NextGenerationEU/PRTR (PRTR-C17.I1),
the ICSC - CentroNazionale di Ricerca in High Performance Computing, Big Data
and Quantum Computing, funded by the European Union NextGenerationEU,
the Comunitat Auton\`oma de les Illes Balears through the Conselleria d'Educaci\'o i Universitats,
the Conselleria d'Innovaci\'o, Universitats, Ci\`encia i Societat Digital de la Generalitat Valenciana and
the CERCA Programme Generalitat de Catalunya, Spain,
the Polish National Agency for Academic Exchange,
the National Science Centre of Poland and the European Union - European Regional
Development Fund;
the Foundation for Polish Science (FNP),
the Polish Ministry of Science and Higher Education,
the Swiss National Science Foundation (SNSF),
the Russian Science Foundation,
the European Commission,
the European Social Funds (ESF),
the European Regional Development Funds (ERDF),
the Royal Society, 
the Scottish Funding Council, 
the Scottish Universities Physics Alliance, 
the Hungarian Scientific Research Fund (OTKA),
the French Lyon Institute of Origins (LIO),
the Belgian Fonds de la Recherche Scientifique (FRS-FNRS), 
Actions de Recherche Concert\'ees (ARC) and
Fonds Wetenschappelijk Onderzoek - Vlaanderen (FWO), Belgium,
the Paris \^{I}le-de-France Region, 
the National Research, Development and Innovation Office of Hungary (NKFIH), 
the National Research Foundation of Korea,
the Natural Sciences and Engineering Research Council of Canada (NSERC),
the Canadian Foundation for Innovation (CFI),
the Brazilian Ministry of Science, Technology, and Innovations,
the International Center for Theoretical Physics South American Institute for Fundamental Research (ICTP-SAIFR), 
the Research Grants Council of Hong Kong,
the National Natural Science Foundation of China (NSFC),
the Israel Science Foundation (ISF),
the US-Israel Binational Science Fund (BSF),
the Leverhulme Trust, 
the Research Corporation,
the National Science and Technology Council (NSTC), Taiwan,
the United States Department of Energy,
and
the Kavli Foundation.
The authors gratefully acknowledge the support of the NSF, STFC, INFN and CNRS for provision of computational resources.

This work was supported by MEXT,
the JSPS Leading-edge Research Infrastructure Program,
JSPS Grant-in-Aid for Specially Promoted Research 26000005,
JSPS Grant-in-Aid for Scientific Research on Innovative Areas 2402: 24103006,
24103005, and 2905: JP17H06358, JP17H06361 and JP17H06364,
JSPS Core-to-Core Program A.\ Advanced Research Networks,
JSPS Grants-in-Aid for Scientific Research (S) 17H06133 and 20H05639,
JSPS Grant-in-Aid for Transformative Research Areas (A) 20A203: JP20H05854,
the joint research program of the Institute for Cosmic Ray Research,
University of Tokyo,
the National Research Foundation (NRF),
the Computing Infrastructure Project of the Global Science experimental Data hub
Center (GSDC) at KISTI,
the Korea Astronomy and Space Science Institute (KASI),
the Ministry of Science and ICT (MSIT) in Korea,
Academia Sinica (AS),
the AS Grid Center (ASGC) and the National Science and Technology Council (NSTC)
in Taiwan under grants including the Science Vanguard Research Program,
the Advanced Technology Center (ATC) of NAOJ,
and the Mechanical Engineering Center of KEK.

Additional acknowledgements for support of individual authors may be found in the following document: \texttt{https://dcc.ligo.org/LIGO-M2300033/public}.
For the purpose of open access, the authors have applied a Creative Commons Attribution (CC BY) license to any Author Accepted Manuscript version arising.
We request that citations to this article use 'A. G. Abac {\it et al.} (LIGO-Virgo-KAGRA Collaboration), ...' or similar phrasing, depending on journal convention.


The MeerKAT telescope is operated by the South African Radio Astronomy Observatory (SARAO), which is a facility of the National Research Foundation, an agency of the Department of Science and Innovation. SARAO acknowledges the ongoing advice and calibration of GPS systems by the National Metrology Institute of South Africa (NMISA) and the time space reference systems department of the Paris Observatory.

Parts of this work were supported by the Australian Research Council Centre of Excellence for Gravitational Wave Discovery (OzGrav) through grants CE170100004 and CE230100016.

MM acknowledges support from the NANOGrav Collaboration's National Science Foundation (NSF) Physics Frontiers Center award numbers 1430284 and 2020265.

C.M.E. acknowledges support from ANID/FONDECYT, grant 1211964. S.G. acknowledges the support of the CNES. W.C.G.H. acknowledges support through grants 80NSSC22K1305 and 80NSSC24K1195 from NASA and Chandra award SAO GO4-25038X.  Chandra grants are issued by the Chandra X-ray Center (CXC), which is operated by the Smithsonian Astrophysical Observatory for and on behalf of NASA under contract NAS8-03060.  This research made use of data obtained from the Chandra Data Archive and the Chandra Source Catalog, and software provided by the Chandra X-ray Center (CXC) in the application packages CIAO and Sherpa.  This work is supported by NASA through the NICER mission and the Astrophysics Explorers Program and uses data and software provided by the High Energy Astrophysics Science Archive Research Center (HEASARC), which is a service of the Astrophysics Science Division at NASA/GSFC and High Energy Astrophysics Division of the Smithsonian Astrophysical Observatory.

S.B.A.F., S.D.P., G.G., F.G., G.E.R. and E.Z. are grateful to the staff of the IAR for technical support and assistance during the observations. The activities at the Instituto Argentino de Radioastronomia (IAR) are supported by the national agency CONICET, the Province of Buenos Aires agency CIC, and the National University of La Plata (UNLP).

E.F. is supported by the National Science Foundation under grant AST-2407399.

Pulsar research at UBC is supported by an NSERC Discovery Grant and the Canadian Institute for Advanced Research.

A.B.P. is a Banting Fellow, a McGill Space Institute~(MSI) Fellow, and a Fonds de Recherche du Quebec -- Nature et Technologies~(FRQNT) postdoctoral fellow.

This document has been given the LIGO-DCC number P2500612.

\software{The 5-vector method is based on the SFDB framework~\citep{Astone:2005fj}, and on the Virgo Rome Snag software~\citep{snag}. Plots are produced using \code{matplotlib}~\citep{2007CSE.....9...90H}. Many pulsar ephemerides are produced with \code{TEMPO}~\citep{2015ascl.soft09002N}, \code{TEMPO2}~\citep{2006MNRAS.369..655H}. The pulsar selection has been handled through \code{CWInPy}~\citep{Pitkin2022} and \code{psrqpy}~\citep{psrqpy}.}

\appendix

\section{UL calculation\label{app:hul}}

We here summarize the UL procedure used to compute the results in Section~\ref{sec:results}. 
The methodology follows the frequentist approach described in Appendix A of~\citet{2017PhRvD..96l2006A}.
Relative to previous narrowband searches, the only methodological modification introduced in this analysis is the inclusion of random 
$\ddot f$ values within the explored parameter space during signal simulations.

ULs are computed independently in each $10^{-4}$~Hz band (following the search setup) using software injections.
For each band, 100 unitary-amplitude simulated signals are generated in the time domain, including the same gaps as in real data, Doppler, and spin-down effects.
After computing the pure-signal five vectors, we rescale them to different amplitudes (starting from $H_0 = 10^{-27}$, well below our theoretical sensitivity), allowing the same simulated signals to be reused efficiently when estimating the detection efficiency as a function of the strain amplitude $H_0$.
The injections are combined with real data at the five-vector level in the frequency domain, and the detection statistic is evaluated. If the computed $\mathcal S$ is above the threshold corresponding to the 1\% false-alarm probability (see Section~\ref{sec:signal_and_method}), we identify the signal as detected. 
If less than 95\% of the signals in a band are recovered, we increase the injected strain by $10^{-28}$ and recompute the detection statistic. 
When the fraction of recovered signals reaches the 95\% CL at $H_{0}^{95\%}$, we compute the upper limit as
    \begin{equation}
        h_{0} ^{95\%} \approx 1.31 \cdot H_{0}^{95\%},
    \end{equation}
where the factor $1.31$ comes from averaging $h_0$ as a function of $H_0$ (i.e., inversion of Equation~\eqref{eq:h0_vs_H0}) over a flat $\cos\iota$ distribution.

\bibliographystyle{aasjournal}
\bibliography{bibliography}

\end{document}